\documentclass[aps,superscriptaddress,pra,twocolumn,footinbib]{revtex4-1}
\usepackage{amsmath,amssymb}
\usepackage{bm}
\usepackage{epsfig}
\usepackage{natbib}
\usepackage{hyperref}
\usepackage{graphicx}
\usepackage{epstopdf}

 \hypersetup{pdfstartview={FitH},pdfpagemode={UseNone},
            colorlinks,linkcolor=red, citecolor=blue, urlcolor=blue,
            bookmarks=true, bookmarksopen=true, pdfnewwindow=true}
             
\usepackage{verbatim} % for multiple-line comment
\usepackage{morefloats}
\usepackage{bibentry}
\usepackage[mathscr]{euscript}

\newcommand{\rmi}{{\rm i}}
\newcommand{\rmd}{{\rm d}}

\newcommand{\e}{{\rm e}}

\renewcommand{\phi}{\varphi}
\begin{document}

\title{
Ring Dirac Solitons  in Nonlinear Topological Lattices
}

\author{\firstname{Alexander N.} \surname{Poddubny}}
\email{poddubny@coherent.ioffe.ru}
\affiliation{Ioffe  Institute, St.~Petersburg 194021, Russia}
\affiliation{Nonlinear Physics Centre, Australian National University, Canberra ACT 2601, Australia}
\author{\firstname{Daria A.} \surname{Smirnova}}
\affiliation{Nonlinear Physics Centre, Australian National University, Canberra ACT 2601, Australia}
\affiliation{Institute of Applied Physics, Russian Academy of Science, Nizhny Novgorod 603950, Russia}

\begin{abstract}
We study solitons of the two-dimensional nonlinear Dirac equation with asymmetric cubic nonlinearity. We show that, with the nonlinearity parameters specifically tuned, a high degree of localization of both spinor components is enabled 
on a ring of certain radius. Such ring Dirac soliton can be viewed as a self-induced nonlinear domain wall and can be implemented in nonlinear photonic graphene lattice with Kerr-like nonlinearities.
Our model could be instructive for understanding localization mechanisms in nonlinear topological systems.
\end{abstract}

\maketitle
%%%%%%%%%%%%%%%%%%%%%
\section{Introduction}\label{sec:intro}
%%%%%%%%%%%%%%%
Dirac equation is a paradigmatic model of  modern physics that describes a plethora of systems ranging from relativistic particles to electrons in graphene, as well as sound~\cite{Rocklin2016}, light~\cite{Amo2017} and cold atoms~\cite{Duca2014} in artificial lattices. Dirac-type models with spin-orbit interactions attract now special attention in condensed matter physics and optics, being applicable to the topological insulators~\cite{bernevig2013,Khanikaev2017,Hadad2018} hosting edge states, immune to disorder. While the linear Dirac equation is now well understood, the physics of the classical nonlinear Dirac model, despite its rich history~\cite{Ranada1983,deSterke1994,Sere2002,Zhdanov2006}, still has a number of open problems. The absence of the rigorous Vakhitov-Kolokolov  criterion\cite{kivshar2003optical} significantly complicates the stability analysis compared to the case of nonlinear Schr\"odinger equation  ~\cite{Bogolubsky1979,Alvarez1983,Kivshar2000,Pelinovsky2014,Comech2014,Saxena2014,Comech2016}. 

There exist a variety of qualitatively different interaction mechanisms introducing nonlinear corrections in the Dirac equation. 
In relativistic field theory the equations have to obey Lorentz invariance that still  leaves Soler~\cite{Soler1970}, Thirring~\cite{Thirring1958,Mikhailov1976}, and Gross-Neveu~\cite{Gross1974}  nonlinear models. Condensed  matter systems, such as  Bose condensates, are not restricted by the Lorentz invariance, which further enriches the family of nonlinear Dirac equations~\cite{Carr2009,Haddad2012,Maraver2017b}. Moreover, the results depend on the dimensionality of the problem. Two-dimensional (2D) Dirac systems deserve special attention since they feature such interesting phenomena  as valley Hall effect and Klein tunneling  and are relatively feasible both in solid state~\cite{Jariwala2016} and optics~\cite{rechtsman2013b,Plotnik2013,Amo2017}. Nonlinear 2D Dirac  equation can be applied to the emerging  field of nonlinear topological photonics~\cite{Segev2013,Lumer2016,Leykam2016,Bardyn2016,Gorlach2017,Zhou2017,Kartashov2017,Solnyshkov2018} aiming for   dynamically tunable  disorder-robust guiding of light.

Recently, a  stable 2D soliton of the nonlinear Dirac equation  has been  found  by Cuevas-Maraver et al.~\cite{Maravero2016,Maraver2017}. The main result of Ref.~\cite{Maravero2016} is the existence and stability of the soliton  in a wide range of parameters for both cubic and quintic nonlinearities. However, the numerical calculation has also revealed an interesting feature in the radial dependence of the spinor amplitude with zero angular momentum ($m=0$). Instead of the expected monotonous decrease from the initial value at $r=0$ to zero at $r=\infty$, the amplitude  features a  weak flat maximum at some radius $r=r^{*}$. Such feature was also previously seen in 1D Soler model and termed as a ``hump''~\cite{Saxena2014}.

In this work, we analyze the generic 2D  nonlinear Dirac equation with asymmetric cubic nonlinearity. We present the physical interpretation of the humped solution as a radial localization of the soliton at the self-induced domain wall, where the band gap in the Dirac equation changes
sign, see Sec.~\ref{sec:numerics}. Such localization is typical for linear topological edge states \cite{bernevig2013} and  hints to an interesting link between nonlinear localization and topological properties. Next, we optimize the nonlinearity parameters in order to achieve the strong degree of localization. We find a stable solution with $m=0$ where both spinor components have prominent maxima at the ring where $r=r^{*}$, and, hence, we term it as a``ring Dirac soliton''. The exact analytical solution of the considered nonlinear Dirac model for the 1D case is given in Sec.~\ref{sec:1D}. The stability and existence analysis for the ring Dirac soliton is presented in Sec.~\ref{sec:stability}.
Finally, in Sec.~\ref{sec:graphene} we propose a scheme  to realize our model in an artificial  photonic graphene lattice with Kerr-like optical nonlinearities.

%%%%%%%%%%%%%%
\section{Ring Dirac Solitons}\label{sec:ring_Dirac}
%%%%%%%%%%%%%%
%%%%%%%%%%%%%%
\subsection{Self-induced domain walls}\label{sec:numerics}
%%%%%%%%%%%%%%

We start with the 2D nonlinear Dirac equation written for the two-component wavefunction $\chi = [\chi_{1},\chi_{2}]$\:,
\begin{align}
\rmi \frac{\partial \chi_{1}}{\partial t}&=E_{1}(\chi_{1},\chi_{2})\chi_{1}-\rmi (\partial_{x}+\rmi \partial _{y})\:, \chi_{2}\label{eq:2D}\\
\rmi \frac{\partial \chi_{2}}{\partial t}&=E_{2}(\chi_{1},\chi_{2})\chi_{2}+\rmi (\partial_{x}-\rmi \partial _{y}) \chi_{1}\:.\nonumber 
\end{align}
Here, 
\begin{align}
E_{1}(\chi_{1},\chi_{2})&=\Delta+a_{1}|\chi_{1}|^{2}+b|\chi_{2}|^{2}\:,\label{eq:E}\\
E_{2}(\chi_{1},\chi_{2})&=-\Delta+a_{2}|\chi_{2}|^{2}+b|\chi_{1}|^{2}\nonumber\
\end{align}
are the positions of the band edges, affected by the cubic nonlinearity, with $\Delta$ being the band gap half-width for zero nonlinearity. In Eqs.~\eqref{eq:E} we consider a general form  of the cubic nonlinearity, that can be asymmetric ($a_{1}\ne a_{2}$). The only restriction is the reciprocity of the cross-coupling nonlinear terms,
$\partial E_{1}/\partial |\psi_{2}|^{2}=\partial E_{2}/\partial |\psi_{1}|^{2}=b$.
We are interested in the solutions with harmonic time dependence and radial symmetry, 
\begin{align}
&\chi_{1}(r,\phi,t)=\psi_{1}(r)\e^{\rmi m\phi-\rmi \omega t},\\ &\chi_{2}(r,\phi,t)= \rmi\psi_{2}(r)\e^{\rmi (m-1)\phi-\rmi \omega t}\:,\label{eq:psichi}
\end{align}
so that Eqs.~\eqref{eq:2D} simplify to the ordinary nonlinear differential equations with respect to the radius $r$:
\begin{align}
\omega\psi_{1}&=E_{1}(\psi_{1},\psi_2)\psi_{1}+\left(\frac{\rmd}{\rmd r}+\frac{1-m}{r}\right)\psi_{2}\label{eq:2Dm}\:,\\
\omega\psi_{2}&=E_{2}(\psi_{1},\psi_2)\psi_{2}-\left(\frac{\rmd}{\rmd r}+\frac{m}{r}\right) \psi_{1}\:.\nonumber
\end{align}
In what follows we focus on the solitons with zero angular momentum, $m=0$, since the high-momentum states
were shown to be unstable~\cite{Maravero2016}.

In order to solve Eqs.~\eqref{eq:2Dm} numerically, we first discretize them using the Chebyshev spectral method, following Refs.~\cite{Trefethen2000spectral,Boyd2001chebyshev}. This scheme provides high precision even for relatively small sets of basis functions. The  singularity at $r=0$ is naturally handled~\cite{Trefethen2000spectral} by looking only for even solutions for $\psi_{1}$ and odd solutions for $\psi_{2}$.  After the discretization is performed, we apply the numerical shooting technique~\cite{Carr2006} to find the solutions, vanishing at $r=\infty$.

\begin{figure}[t]
\includegraphics[width=0.45\textwidth]{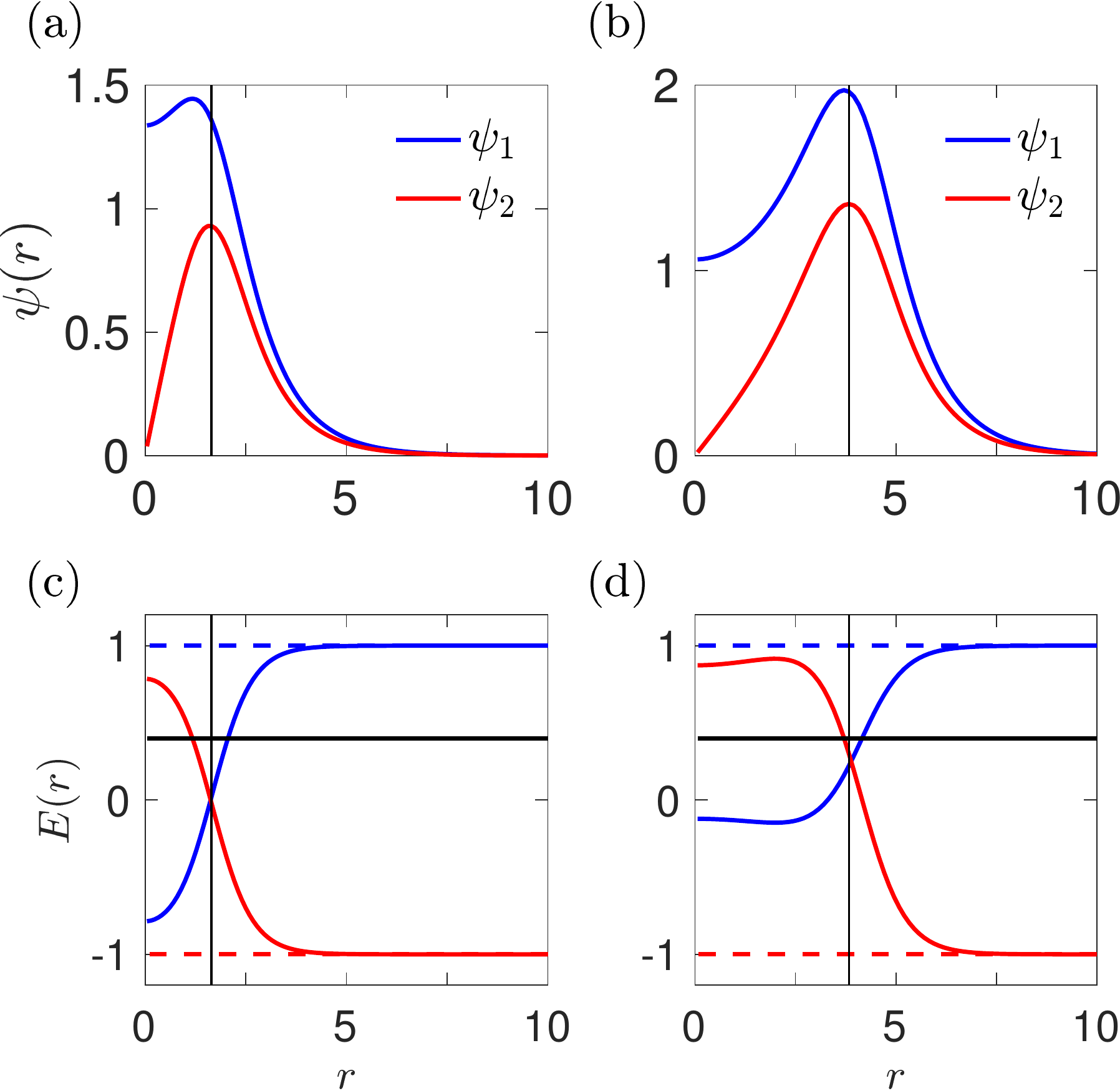}\\
\caption{(a,b) Radial dependence of the spinor components $\psi_{1}$ (blue curve) and 
$\psi_{2}$ (red curve). (c,d) Radial distributions of the {band edges $E_{1}$ and $E_{2}$} for upper and lower bands shown by blue and red curves, respectively. Dashed lines indicate the band edges for vanishing nonlinearity at $E=\pm \Delta =\pm 1$. 
 Thick horizontal line indicates the soliton frequency  $\omega=0.4$.
 Left (a,c) and right (b,d) panels 
correspond to {$\alpha=1$} and {$\alpha=1.67$}, respectively.
Vertical lines indicates the position of the maximum of $\psi_{2}$.
}
\label{fig:1D}
\end{figure}

The results of calculation are presented in Fig.~\ref{fig:1D}. We start the analysis with the case of symmetric nonlinearity,
\begin{equation}
a_{1}=a_{2}=-b=-1\:,\label{eq:NL1}
\end{equation}
corresponding to the 2D Soler model of  Ref.~\cite{Maravero2016}. The obtained spatial distributions of the spinor components  are shown in Fig.~\ref{fig:1D}(a). At the soliton center  the component $\psi_{1}$ has some nonzero value, $r=0$, while the component $\psi_{2}$ is zero. The latter reflects the fact that while the total angular momentum of the soliton is zero, the orbital momentum for the component $\psi_{2}$ is equal to $-1$, see Eq.~\eqref{eq:psichi}. We would like to draw attention to the fact that the both spinor components depend on radius nonmonotonously. They 
have maxima for nonzero value of $r=r^{*}\approx 2$ (indicated by the vertical line) before decaying to zero at $r=\infty$. 
This maximum is practically unresolved for the component $\psi_{1}$, although it becomes more prominent as the soliton frequency decreases. To the best of our knowledge, the physical interpretation of this maximum has so far not been presented.
%%%%%%%%%%%%%%%%%%%%%%
%%%%%%%%%%%%%%%%%%%%%%

The origin of such hump can be understood if  one examines the radial dependences of the band edges, shown 
of the band edge positions $E_{1,2}(r)$, presented in Fig.~\ref{fig:1D}(c). At large radii where the nonlinear corrections vanish they tend to the linear band gap edges $\pm \Delta$. The soliton frequency $\omega$ lies inside the band gap, in agreement with the exponential decay of the amplitude. When the radius decreases and the soliton components grow, the nonlinearity induces the crossing of band edges. Their positions at $r=0$ are swapped as compared to the case $r\to\infty$. The crossing point matches the maxima of $\psi_{1,2}$. 
Spatial localization of eigenmodes at the band crossing in the Dirac spectrum is very intuitive in the linear regime and lies at the heart of the topological edge states \cite{Jackiw1976,Shen2013,bernevig2013}. In this case the domains with opposite band orders are described by certain integer topological invariants and the domain wall hosts a localized state. The possibility of  topological characterization of nonlinear edge states is so far unclear. However,  the general phenomenon behind the localization in topological insulators and the maximum of spinor components in considered nonlinear Dirac equation is the same band crossing. As such, one can interpret the Dirac soliton in 
Fig.~\ref{fig:1D}(a) as a self-induced domain wall between the circular core with $E_{1}<E_{2}$ and the outlying space with $E_{1}>E_{2}$.

The maximum in Fig.~\ref{fig:1D}(a) for $\psi_{1}$ is not particularly prominent. It becomes sharper as the frequency decreases, however, at low frequencies $|\omega|\lesssim 0.14$ the soliton loses stability with respect to the radial perturbations with the angular momentum $m'=2$~\cite{Maravero2016}. Hence, it is an interesting question whether one can tune the nonlinearity to make the maximum more prominent while simultaneously  keeping the soliton frequency the same and the soliton stable.

The signature of the self-induced domain wall would be a realization of the band crossing condition
\begin{equation}
E_{1}(\chi^{(1)})=\omega,\quad E_{2}(\chi^{(2)})=\omega \label{eq:condition}
\end{equation}
at the frequency $\omega$ for the same value $\chi^{(1)}=\chi^{(2)}$. One can see from Fig.~\ref{fig:1D}(c) that 
the bands cross at $E_{1}=E_{2}=0$ relatively far below the soliton frequency $\omega=0.4$, so that the radial localization is weak.
 The spinors  $\chi^{(1)}$ and $\chi^{(2)}$ in Eq.~\eqref{eq:condition} can be estimated by extending  the evanescent linear asymptotic solutions of Eqs.~\eqref{eq:2Dm} with the ratio of spinor components  $\chi^{(1,2)}_{1}/\chi^{(1,2)}_{2}=\sqrt{(1+\omega)/(1-\omega)}$ from $r\to \infty$ to smaller values of $r$. Substituting this ratio into Eq.~\eqref{eq:condition} we find 
\begin{equation}
\xi\equiv \left(\frac{\chi^{(1)}_{1}}{\chi^{(2)}_{1}}\right)^{2}=-\frac{1-\omega}{1+\omega}\frac{a_{2}(1-\omega)+b(1+\omega)}{a_{1}(1+\omega)+b(1-\omega)}\:.\label{eq:chi}
\end{equation} 
The ideal situation, when both bands cross the frequency $\omega$ at the same time, corresponds to $\xi=1$. Our
recipe to improve the radial localization at a given frequency is to adjust the nonlinearity by  making $\xi$ closer to unity. Inspection of Eq.~\eqref{eq:chi} shows that this condition is further simplified when the symmetry relation 
$a_{1}/b=b/a_{2}$ holds, so that
\begin{equation}
a_{1}=-1, \quad a_{2}=-\alpha^{2},\quad b=\alpha\:,\label{eq:NL2}
\end{equation}
and 
\begin{equation}
\xi= \alpha\frac{1-\omega}{1+\omega}\:.\label{eq:chi2}
\end{equation}
Hence, one can expect that for 
$\xi$ becomes closer to unity and the radial localization is improved  
when $\alpha>1$ for $\omega>0$ and  $\alpha<1$ for $\omega<0$.

\begin{figure}[t!]
\includegraphics[width=0.45\textwidth]{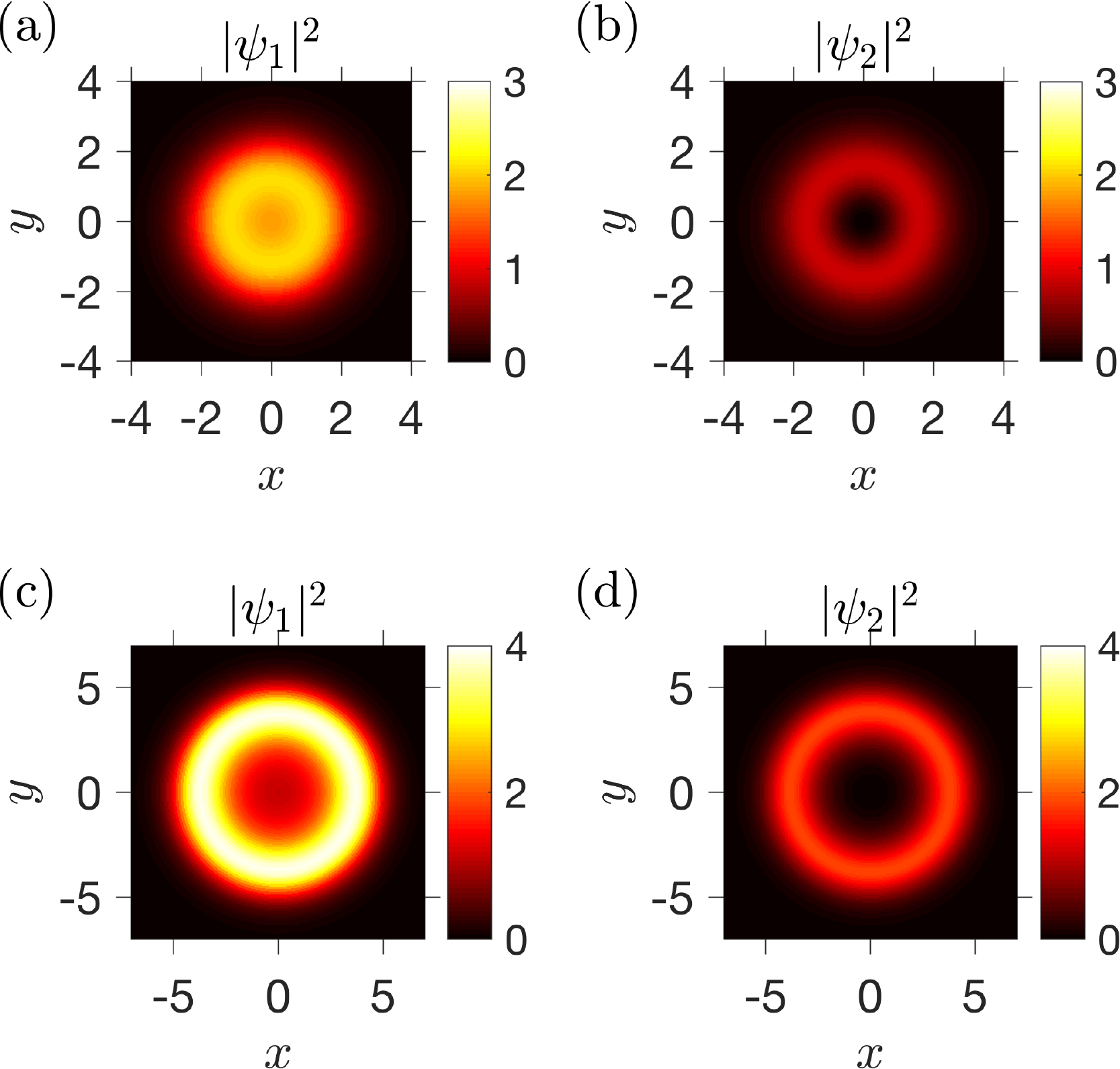}\\
\caption{Spatial distributions of the  soliton spinor components  $|\chi_{1}|^{2}$ (a,c) and 
$|\chi_{2}|^{2}$ (b,d), corresponding to the parameters of Fig.~\ref{fig:1D}.
Panels (a,b) and (c,d) were calculated for  {$\alpha=1$} and {$\alpha=1.67$}, respectively.}
\label{fig:2D}
\end{figure}

In order to test this analytical prediction we show in  Fig.~\ref{fig:1D}(c,d)  the results for asymmetric nonlinearity coefficients satisfying Eq.~\eqref{eq:NL2} with $\alpha=1.67$. While the  structure of the solution remains formally the same as in Figs.~\ref{fig:1D}(a,b), where $\alpha=1$, it now features a prominent maximum for both $\psi_{1}$ and $\psi_{2}$ components, clearly pinned to the nonlinear band edge crossing point.
The distinction between the symmetric and asymmetric cases is most clearly seen in the two-dimensional maps of the soliton, $|\chi_{1,2}(x,y)^{2}|$, shown in Fig.~\ref{fig:2D}.  A ring at $r\approx 4$ is clearly visible for both spinor soliton components  in the asymmetric case [Fig.~\ref{fig:2D}(b,d)] while it is practically unresolved in  Fig.~\ref{fig:2D}(a) in the symmetric case.  Various solitonic and vortex solutions of the Dirac equation are known, see e.g.  a detailed analysis in Ref.~\cite{Haddad2015e}. However, to the best of our knowledge, the $m=0$ Dirac solitons with rings for both spinor components have not been analyzed before.

%%%%%%%%%%%%%%%%%%%%%%
%%%%%%%%%%%%%%%%%%%%%%

%%%%%%%%%%%%%%%%%%%%%%
\subsection{Exact solution in 1D}\label{sec:1D}
%%%%%%%%%%%%%%%%%%%%%
In order to gain deeper insight in the properties of the ring Dirac solitons we consider a simplified one-dimensional problem instead of the cylindrically symmetric two-dimensional one. The 1D problem with symmetric nonlinearity is known to admit  exact analytical solutions \cite{Khawaja2014}. It turns out that  the nonlinearity of type Eq.~\eqref{eq:NL2} admits them as well. In the 1D case the derivative over $x$ in Eqs.~\eqref{eq:2D} can be suppressed and they reduce to
\begin{align}  \label{eq:1D}
\omega \chi_1 &= E_{1}(\chi_{1},\chi_{2}) \chi_1 +\dfrac{\rmd \chi_2}{\rmd y}  \:,\\
\omega \chi_2 & = E_{2}(\chi_{1},\chi_{2}) \chi_2 -\dfrac{\rmd \chi_1}{\rmd y}  \nonumber\:.
\end{align}
Equations~\eqref{eq:1D} have a Hamiltonian structure 
\[
\frac{\rmd\chi_{1}}{\rmd y}=\frac{\partial H}{\partial \chi_{2}},\quad 
\frac{\rmd\chi_{2}}{\rmd y}=-\frac{\partial H}{\partial \chi_{1}}
\]
with the Hamiltonian 
\begin{equation}  \label{eq:3}
H (\chi_{1},\chi_{2})= \dfrac{\omega -\Delta}{2} \chi_1^2 + \dfrac{\Delta+\omega}{2} \chi_2^2 + \dfrac{1}{4} (\chi_1^2 - \alpha \chi_2^2)^2 \:.
\end{equation}
For localized soliton solutions $\chi_{1,2} (\infty) \rightarrow 0$, and, hence, $H=0$. We use the substitution 
\begin{equation}  \label{eq:4}
  \chi_1 (y)  = A(y) \cos{\varphi(y)} \:,\quad 
 \chi_2 (y) = A(y) \sin{\varphi(y)} \:,
\end{equation}
so that Eq.~\eqref{eq:3} yields
\begin{equation}  \label{eq:5}
A^2 = \dfrac{2(\Delta\cos{2\varphi} - E)}{ (\cos^2{\varphi}  - \alpha \sin^2{\varphi})^2} \:.
\end{equation}
Substituting Eq.~\eqref{eq:4} into Eq.~\eqref{eq:1D}  we obtain an ordinary differential equation  for $\varphi$  that turns out to be $\alpha$-independent
\begin{equation}  \label{eq:6}
\dfrac{\rmd \varphi}{\rmd y } = - E + \Delta \cos {2\varphi} \:,
\end{equation}
and has the solution 
\begin{equation}
\varphi =\arctan\left[  \sqrt{\dfrac{\Delta - \omega}{\Delta+\omega}} \tanh \left(\sqrt{\Delta^2 - \omega^2} y \right)\right]\:.
\label{eq:phi}
\end{equation}
The final expressions for $\chi_{1,2}$ are obtained by substituting Eq.~\eqref{eq:phi} and Eq.~\eqref{eq:5} into Eq.~\eqref{eq:4}.

%%%%%%%%%%%%%%%%%%%%%
\subsection{Existence and Stability Analysis}\label{sec:stability}
%%%%%%%%%%%%%%%%%%%%%
%%%%%%%%%%%%%%
\begin{figure}[t!]
\includegraphics[width=0.5\textwidth]{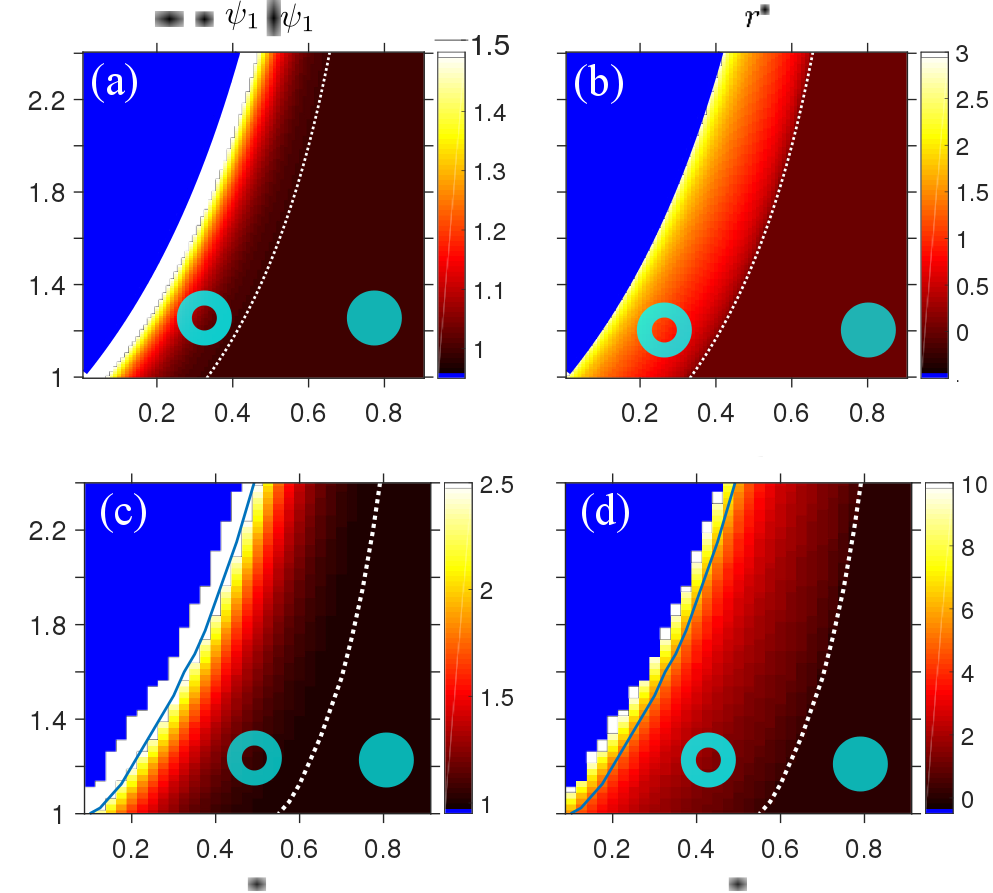}
\caption{Shape of the soliton depending on its frequency $\omega$ and the nonlinearity asymmetry $\alpha$.
(a,c):  ratio of the maximum of the component $\chi_{1}$ at the point $r^{*}$ to its value at $r=0$. (b,d):  position of the soliton maximum $r^{*}$. Panels (a,b) have been calculated analytically for the 1D soliton solution Eq.~\eqref{eq:4},  panels (c,d) correspond to  numerical calculation in 2D. 
Blue line in  (c,d) bounds the region  where the soliton is spectrally unstable and in the blue region soliton does not exist. Dotted white line bounds the  region where the soliton has a ring shape, i.e. $r^{*}>0$. Calculation has been performed for $\Delta=1$.
}
\label{fig:sweep}
\end{figure}
%%%%%%%%%%%%%%%%%%%%%
In order to determine the spectral stability of the soliton solution $\chi^{(0)}$, we perform a standard linearization procedure. The solution is sought at the frequency $\omega'$ in the form 
\begin{equation}
\chi^{(1)}=\chi^{(0)} \e^{-\rmi \omega t}+U \e^{-\rmi \omega' t}+V\e^{+(\rmi \omega'{}^{*}-2\rmi \omega) t}\:,
\end{equation}
where $\chi^{(0)}$ is the unperturbed soliton profile, and the spinors $u$ and $v$ have the form
\begin{equation} 
U=\begin{pmatrix}
u_{1}\e^{\rmi (m+m')\phi}\\\rmi u_{2}\e^{\rmi (m+m'-1)\phi}
\end{pmatrix},\quad 
V=\begin{pmatrix}
v_{1}\e^{\rmi (m-m')\phi}\\\rmi v_{2}\e^{\rmi (m-m'-1)\phi}
\end{pmatrix}\:,
\end{equation}
and $m'$ is the angular momentum of the perturbation.
The resulting system of linear equations for the spinors $u=[u_{1},u_{2}]^{T}$, 
$v=[v_{1},v_{2}]^{T}$
reads
\begin{equation}
\omega'\begin{pmatrix}u\\v
\end{pmatrix}=
\begin{pmatrix}
H_{+}&H_{0}\\-H_{0}&2\omega-H_{-}
\end{pmatrix}\begin{pmatrix}u\\v
\end{pmatrix}
\end{equation}
with 
\begin{align}
H_{\pm}&=\begin{pmatrix}
\Delta+2a_{1}\psi_{1}^{2}+b\psi_{2}^{2} &D_{+,m\pm m'}+b\psi_{1}\psi_{2}\\
-D_{-,m\pm m'}+b\psi_{1}\psi_{2}&-\Delta+2a_{2}\psi_{2}^{2}+b\psi_{1}^{2}
\end{pmatrix}\:,\nonumber\\\nonumber
H_{0}&=\begin{pmatrix}
a_{1}\psi_{1}^{2}&b\psi_{1}\psi_{2}\\b\psi_{1}\psi_{2}&a_{2}\psi_{2}^{2}
\end{pmatrix}\:,
\end{align}
where the differential operators are
\begin{equation*}
D_{+,m}=\frac{\partial}{\partial r}+\frac{1-m}{r},\quad D_{-,m}=\frac{\partial}{\partial r}+\frac{m}{r}\:.
\end{equation*}

The calculated stability and existence diagram for the ring soliton  depending on its frequency and 
nonlinearity asymmetry $\alpha$ is presented in Fig.~\ref{fig:sweep}.
Figures~\ref{fig:sweep}(a,c) show the relative amplitude of the maximum, $\psi_{1}(r^{*})/\psi_{1}(0)$, while Figs.~\ref{fig:sweep}(b,d) present the maximum position. 
 We start the analysis from the exact 1D solution, obtained in the previous section. The soliton exists and has a double-hump profile with maxima at $y=\pm r^{*}$ in the region where 
\begin{equation}
\frac{1}{2}\frac{1+\omega}{1-\omega}<\alpha<\frac{1+\omega}{1-\omega}\:.
\end{equation} 
For lower values of $\alpha$ the hump disappears and the maximum is in the coordinate origin, i.e. $r^{*}=0$. The  boundary of the corresponding region is indicated by a black dashed curve. For $\alpha$ larger than $(1+\omega)/(1-\omega)$ the solution does not exist. The main result of the calculation is the possibility to adjust the soliton shape by controlling the nonlinearity. For a given  soliton frequency, the increase of the $\alpha$ parameter transforms the ordinary soliton to the double-hump one and enhances the maximum for $\psi_{1}^{*}$ component, in agreement with the calculation in Fig.~\ref{fig:1D}.

Figures~\ref{fig:sweep}(c,d) show the phase diagram for the 2D ring solution solution. It has qualitatively the same structure as the one in the 1D case. However, the ring soliton region where $r^*>0$ (bounded by dotted white curve) is considerably wider in the 2D case than in the 1D case, cf. Fig.~\ref{fig:sweep}(b) and Fig.~\ref{fig:sweep}(d).
Blue line indicates the stability region. Increase of the nonlinearity asymmetry first makes the ring soliton unstable and than it ceases to exist. Both for symmetric and asymmetric nonlinearity the instability develops first for the perturbation with $m'=2$.

%%%%%%%%%%%%%%%%%%%%%
\section{Implementation in nonlinear lattices}\label{sec:graphene}
%%%%%%%%%%%%%%%%%%%%%
%%%%%%%%%%%%%%%%%%%%%
\begin{figure}[t]
\centering\includegraphics[width=0.45\textwidth]{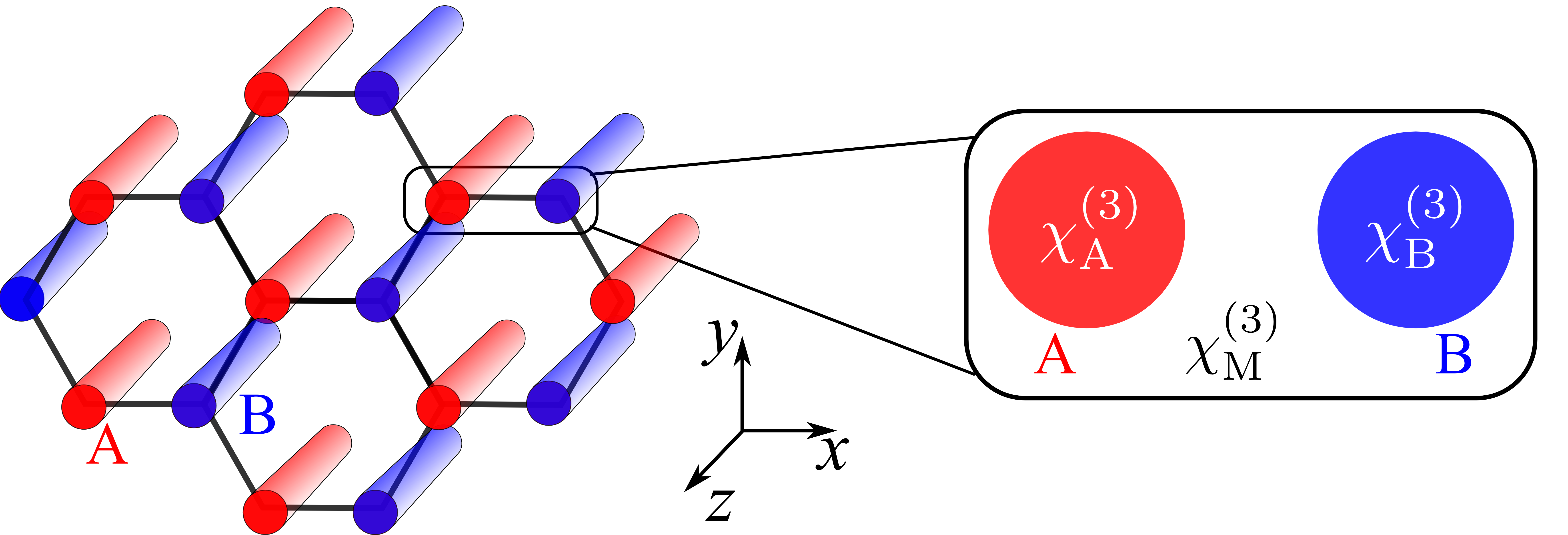}
\caption{Schematic of photonic graphene structure with  waveguides having the opposite sign  of 
Kerr nonlinearity than the matrix, $\chi^{(3)}_{\rm A,B}\chi^{(3)}_{\rm M}<0$.}
\label{fig:Graphene}
\end{figure}
%%%%%%%%%%%%%%%%%%%%%
Here, we consider a model example of nonlinear photonic graphene
based on an array of waveguides arranged in a honeycomb lattice, see Fig.~\ref{fig:Graphene}. In the linear optical regime this waveguide platform  has been already used to demonstrate topological edge states of light \cite{Plotnik2013,rechtsman2013,rechtsman2013b}.
Our suggestion is to embed waveguides with defocusing Kerr nonlinearity ($\chi^{(3)}<0$)  in a focusing nonlinear matrix with $\chi^{(3)}>0$.
The tight binding equations governing the propagation of light in this system along $z$ direction  have the following form:
\begin{align}\label{eq:TB1}
-\rmi \frac{\partial \psi_{A,j}}{\partial z}&=E_{A,j}\psi_{A,j}+t\sum\limits_{\langle j,j'\rangle }\psi_{B,j'}\:,\\\nonumber
-\rmi \frac{\partial \psi_{B,j}}{\partial z}&=E_{B,j}\psi_{B,j}+t\sum\limits_{\langle j,j'\rangle }\psi_{A,j'}\:,\nonumber\\\nonumber
E_{A,j}&=\Delta+a_{1}|\psi_{A,j}|^{2}+\frac{b}{3}\sum\limits_{\langle j,j'\rangle }|\psi_{B,j'}|^{2}\:,\\\nonumber
E_{B,j}&=-\Delta+a_{2}|\psi_{B,j}|^{2}+\frac{b}{3}\sum\limits_{\langle j,j'\rangle }|\psi_{A,j'}|^{2}\:.\nonumber
\end{align}
Here, $\psi_{A,j}$ and $\psi_{B,j}$ are the smooth envelopes of electric field $\mathcal E$ at the sites $A$ and $B$ of the unit cell $j$; $\mathcal E(z,j,t)\propto \e^{\rmi k_{z}z-\rmi \omega t}\psi_{j}(z)$, where $\omega$ is the frequency and $k_{z}$ is the wave vector along the waveguide axis. The parameter  $t$ describes  the tunneling, $\sum_{\langle j,j'\rangle }$ denotes the sum over nearest neighbors of the cell $j$, 
 $\Delta$  is  the  detuning of modes of the waveguides $A$ and $B$ and can be controlled by varying the waveguide radiuses.
We consider only the linear tunneling terms but take into account both local ($\propto a_{1,2}$) and nonlocal ($\propto b$) Kerr nonlinearities.  
 Discrete nonlinear lattices manifest  interesting phenomena such as self-induced gap soliton formation ~\cite{Kivshar1993,Kivshar1994b}. In case of  Dirac systems with symmetric nonlinearity the effects of discreteness  were recently analyzed in Ref.~\cite{Maraver2017b}. For the purpose of this work it is sufficient to apply the smooth envelope approximation near the Dirac points. We start with   the solution of Eqs.~\eqref{eq:TB1} in the Bloch form, 
\begin{equation}
\psi_{A,j}=\psi_{A}\e^{\rmi \bm k\cdot \bm r_{j}},\quad 
\psi_{B,j}=\psi_{B}\e^{\rmi \bm k\cdot \bm r_{j}}\:.
\end{equation}
which results in 
\begin{align}
-\rmi \frac{\partial \psi_{A}}{\partial z}&=\left(\Delta+a_{1}|\psi_{A_{j}}|^{2}+b|\psi_{B}|^{2}\right )\psi_{A}+g_{\bm k}\psi_{B}\\
-\rmi \frac{\partial \psi_{B}}{\partial z}&=\left(-\Delta+a_{2}|\psi_{B_{j}}|^{2}+b|\psi_{A}|^{2}\right)\psi_{B}+g^{*}_{\bm k}\psi_{A}\nonumber
\end{align}
where $g_{\bm k}=t[\e^{\rmi k_{x}d}+2\e^{-\rmi k_{x}d/2}\cos(k_{y}d)]$ and $d$ is the distance between nearest neighbors.  Applying the $\bm k\cdot \bm p$ expansion near the Dirac points 
\begin{equation}
\kappa_{x}=k_{x},\quad \kappa_{y}=k_{y}\mp\frac{4\pi\sqrt{3}}{9d}
\end{equation}
we obtain the following equations
\begin{align}
-\rmi \frac{\partial \psi_{A}}{\partial z}&=\left(\Delta+a_{1}|\psi_{A}|^{2}+b|\psi_{B}|^{2}\right )\psi_{A}+v(\rmi \kappa_{x}\mp \kappa_{y})\psi_{B}\:,\label{eq:graphene}\\
-\rmi \frac{\partial \psi_{B}}{\partial z}&=\left(-\Delta+a_{2}|\psi_{B}|^{2}+b|\psi_{A}|^{2}\right)\psi_{B}-v(\rmi \kappa_{x}\pm \kappa_{y})\psi_{A}\:,\nonumber
\end{align}
where $v=3t/2$ and the sign $\pm$ corresponds to two different valleys.
In the smooth envelope approximation, when $\bm \kappa$ is replaced by $-\rmi \bm \nabla$, equations \eqref{eq:graphene} are equivalent to the nonlinear Dirac equations Eqs.~\eqref{eq:2D},\eqref{eq:E}  for each of the two valleys. The condition 
$a_{1,2}b <0$, required for the ring Dirac soliton formation, can be realized by controlling the nonlinear response of the waveguides and of the matrix.
Namely,  defocusing nonlinearity in each of the waveguides, $\chi^{(3)}_{\rm A,B}<0$, decreases the wavevector along $z$, hence, $a_{1},a_{2}<0$. On the other hand, the nonlocal Kerr term results from the spatial overlap between the evanescent tails
of the electric fields from neighboring waveguides, penetrating in the matrix. Thus, when the matrix has focusing Kerr nonlinear response with $\chi_{\rm M}^{(3)}>0$, increase of the field intensity in the given waveguide will increase the wave vector for  its neighbors, hence, $b>0$. 

Implementation of the cubic nonlinearity with modulated sign seems to be a challenging but feasible task, see a detailed review in Ref.~\cite{Kartashov2011}. For example, one could consider an optofluidic platform~\cite{Kivshar2009,Giessen2010} with photonic crystal fibers filled by liquids. Strong fast focusing nonlinearity is   available in  chalcogenide glass fibers ~\cite{Coulombier2010}. Chalcogenide waveguides with   high  nonlinear refractive index  $n_{2}\sim  4\times 10^{-14}$~cm$^{2}/$W~\cite{Asobe1992}  have enabled demonstration of  self-focusing   for cm-long samples at the $1.5~\mu$m wavelength~\cite{Chauvet2009}.
The main difficulty would be to balance the    focusing nonlinearity of the fiber material with slow but strong thermal defocusing nonlinearity of the liquids, where values up to $n_{2}=-1.6\times 10^{-5}~$cm$^{2}$/W were reported~\cite{Smith2014}.  
In the pulsed regime this could be done by controlling the repetition rate. Alternatively, one could consider solutions where the thermal defocusing nonlinearity is relatively weak, $n_{2}\sim - 10^{-13}$~cm$^{2}/$W~\cite{Cheung1994}.
In addition to the photonic crystal waveguide platform, similar effects could be potentially realized for excitonic polaritons in arrays of coupled micropillars with embedded quantum wells~\cite{Amo2017} and for cold atom lattices \cite{Duca2014} with nonlinearity controlled by Feshbash resonances~\cite{Roberts2009}.

%%%%%%%%%%%%%%%%%%%%%%%%%%%
\section{Conclusion}\label{sec:conclusions}
%%%%%%%%%%%%%%%%%%%%%%%%%%%%%%%%%%%%%
To summarize, we have theoretically studied rotationally symmetric soliton solutions of two-dimensional nonlinear Dirac equation with asymmetric cubic nonlinearity. It has been demonstrated that the radial distribution of the soliton amplitude strongly depends on the nonlinearity parameters. By adjusting the nonlinearity we were able to find a spectrally stable ring Dirac  solution where both spinor components of the soliton have maxima at a certain radius.  The radial localization is explained by the crossing of the band gap edges in the Dirac equation, induced by the nonlinearity. Hence, the obtained solution can be interpreted as a self-induced nonlinear domain wall. 
A condition relating the  radial localization degree to the nonlinearity parameters has been derived.
We present a conceptual scheme to realize ring Dirac soliton in a nonlinear photonic graphene lattice with defocusing  waveguides in the  focusing matrix or vice versa. 

Remarkably, the formation of solitons at the self-induced  domain walls is intuitively quite similar to localization of  edge states at the boundary between domains with distinct topological invariants.  However, despite detailed  studies of edge modes in nonlinear lattices~\cite{Kartashov2011}, no topological classification, comparable with the established classification of topological insulators~\cite{Ludwig2010}, is available for nonlinear edge states. Even though generalizations of the Berry phase in nonlinear systems are known ~\cite{Liu2010}, the possibility of rigorous nonlinear bulk-boundary correspondence remains an open question. We believe that our model can be applicable to a variety of nonlinear topological systems spanning from  photonics to  cold atom lattices   and could provide important insights for further developments in the nonlinear topological physics.

\acknowledgements
The authors thank Yu.S.~Kivshar, E.A.~Ostrovskaya, A.V.~Yulin, I.V. Iorsh, D.R. Gulevich,  and A.A.~Sukhorukov for stimulating discussions.
A.N.P. acknowledges the support of the ``Basis'' Foundation.
D.A.S. acknowledges the support of the Russian Foundation for Basic Research (Grant No. 18-02-00381).

%see https://tex.stackexchange.com/questions/254423/applying-longbibliography-option-without-using-%documentclassrevtex4-1
%%adding custom command to revtex to have initialed author names AND titles
%\nocite{apsrev41Control}
%\bibliographystyle{apsrev4}
%\bibliography{titleon,Dirac}

\end{document}